\UseRawInputEncoding
\documentclass[aps,showpacs,prep,graphics]{revtex4}

\usepackage{amssymb}
\usepackage[dvips]{graphicx}
\usepackage{amsmath}
\usepackage{bm}
\usepackage{epsfig}
\usepackage{graphicx}
\usepackage{caption}
\usepackage{wrapfig}
\usepackage{color}
\usepackage{subfig}

\begin{document}

\title{Power spectrum of scalar fluctuations of the metric during the formation of a scalar black hole in inflation}

\author{ $^{1}$ Jos\'e Edgar Madriz Aguilar,  $^{2}$ J. O. Valle, $^{1}$ M. Montes , $^{3}$ C. Romero \bigskip
\thanks{E-mail address: mariana.montnav@gmail.com} }
\affiliation{$^{1}$ Departamento de Matem\'aticas, Centro Universitario de Ciencias Exactas e ingenier\'{i}as (CUCEI),
Universidad de Guadalajara (UdG), Av. Revoluci\'on 1500 S.R. 44430, Guadalajara, Jalisco, M\'exico,  \\
and\\
$^{2}$ Centro Universitario de los Valles \\
Carretera Guadalajara-Ameca Km 45.5,\\
C. P. 46600, Ameca, Jalisco, M\'exico,\\
and\\
$^{3}$ Departamento de F\'{i}sica\\
Universidade Federal da Paraíba, Caixa Postal 
 5008,\\
 58059-970, Jo\~{a}o Pessoa, PB, Brazil\\
 \\ 
E-mail:  jose.madriz@academicos.udg.mx, juan.valle@alumnos.udg.mx, mariana.montes@academicos.udg.mx, cromero@fisica.ufpb.br }

\begin{abstract}

 In this paper we use the collapse metric obtained by Carneiro and Fabris to calculate the power spectrum associated to gauge invariant fluctuations of the metric during the formation of a primordial scalar black hole at the end of inflation. We assume that local perturbations in the vacuum energy density can collapse by means of the collapse of the local inflaton field, generating a sin\-gu\-la\-ri\-ty as shown by Carneiro and Fabris. Employing a representative term of the series expansion of the scale factor we obtain a nearly scale invariant power spectrum for a second order approximation of the scale factor. Thus, the spectrum for scalar black holes obtained can be considered as a correction of the power spectrum of the primordial quantum fluctuations of the inflaton field. 
\end{abstract}

\maketitle

\vskip .5cm
 Primordial scalar black holes, inflation, scalar fluctuations of the metric.

\section{INTRODUCTION}

Due to their spectrum of masses black holes are nowadays considered as good candidates for dark matter. Some are originated due to  gravitatational collapse of stars. However, there are  others, called primordial black holes (PBHs), that are believed to have been formed in the early universe \cite{RI1,RI2}. The two possible mass ranges for PBHs to significantly contribute to dark matter are: the asteriod mass region between $10^{-16}\,M_{\odot}$, and $10^{-11}\,M_{\odot}$, and the solar-mass region that is only slightly constrained by microlensing surveys \cite{RI3}. \\

Many models concerning about the origin of PBHs are baesd on first and second order transitions \cite{RI4,RI5}. However, we can find in the literature several other proposals like for example the collapse of density perturbations generated during the inflationary period \cite{RI6,RI7}, the dynamics of scalar condensantes \cite{RI8}, and during the collapse of topological defects \cite{RI9}, among others. \\

On the other hand,  Carneiro and Fabris have shown that the gravitational collapse of a minimally coupled scalar field collapses to a singularity when the energy-momentum tensor of the scalar field is separated into a pressureless matter and a vacuum parts \cite{Carneiro}. Hence, if we consider this mechanism at the end of inflation we can interpret that local fluctuations in the vacuum energy density can collapse to form a type PBHs called scalar black holes. \\

In this paper we use the collapse metric obtained by Carneiro and Fabris to calculate the power spectrum generated during the collapse of a local perturbation in the vacuum energy density at the end of inflation. With this in mind the paper was organized as follows.  In section I we give a brief introduction. In section II we explain the formalism for non-perturbative gauge invariant fluctuations of the metric during the collapse. In section III we treat the case of scalar fluctuations of small amplitude as a weak field limit. In section IV we calculate the power spectrum to the gauge invariant fluctuations of the metric generated during the formation of a scalar black hole. Finally section V is left for some final comments.

\section{The formalism }

Let us start considering the energy-momentum tensor of the inflaton scalar field in the usual form
\begin{equation}\label{eq1}
    T_{\alpha\beta}=\varphi_{,\alpha}\varphi_{,\beta}-g_{\alpha\beta}\left(\frac{1}{2}\varphi^{,\sigma}\varphi_{,\sigma}+V(\varphi)\right),
\end{equation}
with $V(\varphi)$ being the potential associated to the inflaton scalar field and $g_{\mu\nu}$ the space-time metric. Now, following the procedure exposed in \cite{Carneiro}, we locally consider the decomposition of \eqref{eq1} as
\begin{equation}\label{eq2}
T_{\alpha\beta}=T^{(m)}_{\alpha\beta}+T^{(\Lambda)}_{\alpha\beta},
\end{equation}
where the pressureless and vacuum componentes are given respectively by
\begin{eqnarray}\label{eq3}
    T^{(m)}_{\alpha\beta}&=&\varphi_{,\alpha}\varphi_{,\beta},\\
    \label{eqp3}
    T^{(\Lambda)}_{\alpha\beta}&=&\left(V(\varphi)-\frac{1}{2}\varphi^{,\sigma}\varphi_{,\sigma}\right)g_{\alpha\beta}=\Lambda g_{\alpha\beta}.
\end{eqnarray}
Furthemore, the density of matter is defined as $\rho=g^{\alpha\beta}T^{(m)}_{\alpha\beta}$, and the 4-velocity is given by
\begin{equation}\label{eq5}
    u^\mu=\frac{\partial^\mu\varphi}{\sqrt{\partial^\alpha\varphi\partial_\alpha\varphi}}.
\end{equation}
Using this decomposition, a massive scalar field can generate a spherical gravitational collapse leading to the formation of scalar black holes \cite{Carneiro}. 
In co-moving coordinates, the collapsing metric within the local spacetime region with spherical symmetry is given by \cite{Carneiro} 
\begin{equation}\label{eq6}
    ds^2=dt^2-a^2(t)\left[\frac{F'^2(r)}{9F^{4/3}(r)}dr^2+F^{2/3}(r)\left(d\theta^2+\sin^2(\theta) d\phi^2 \right)\right],
\end{equation}
where $F(r)$ is an arbitrary function of $r$ with the condition $F(0)=0$, and the scale factor reads
\begin{equation}\label{eq7}
    a(t)=\left(\frac{3}{\Lambda}\right)^{1/3}\sinh^{2/3}\left[\frac{\sqrt{3\Lambda}}{2}(t-t_0)\right],
\end{equation}
where $\Lambda$ is the vacuum energy density given by \eqref{eqp3}, and $t_0$ is the time at which the collapse begins. If we examine the scale factor (\ref{eq7}) we notice that when the time $t$ coincides with the initial collapse time $t=t_0$, the scale factor $a_0$ is zero, this could cause singularity problems in the formalism of scalar fluctuations of the metric. So, to alleviate the inconvenient we introduce the parameter $\epsilon$ defined as
\begin{equation}\label{eq8}
    \epsilon=\frac{r_g}{r_{H}},
\end{equation}
where $r_g$ is the gravitational radius of a formed scalar field black hole, and $r_{H}$ is the Hubble radius by the time when the inflationary modes re-enter the horizon. It is not  difficult to see that $r_{H}\gg r_g$ thus $\epsilon\ll 1$. Thus $a(t_0+\epsilon t_0)\simeq a(t_0)$ and we can avoid possible singularity problems. \\

Now, in order to study gauge invariant scalar fluctuations of the metric during the gravitational collapse, we will use the non-perturbative formalism employed in \cite{Anabitarte}. Thus, the perturbed line element can be written in the form
\begin{equation}\label{eq9}
    ds^2=e^{2\psi} dt^2-a^2(t)e^{-2\psi}\left[\frac{F'^2(r)}{9F^{4/3}(r)}dr^2+F^{2/3}(r)(d\theta^2+\sin^2(\theta)d\phi^2)\right],
\end{equation}
where $\psi(t,r,\theta,\phi)$ is a metric function that describes the gauge-invariant scalar fluctuations of the metric in a non-perturbative way. With the help of \eqref{eq9}, the Ricci scalar reads
\begin{equation}\label{eq10}
    R=\frac{2e^{2\psi}}{F^{2/3}a^2}\left[\bar{M}(\psi)-\bar{N}(\psi)+3\left(4\frac{F}{F'}-3\frac{F''F^2}{F'^3}\right)\frac{\partial\psi}{\partial r}+\cot(\theta)\frac{\partial\psi}{\partial \theta}\right]+6e^{-2\psi}\left[3\dot{\psi}^2-5\dot{\psi}H-\ddot{\psi}+H^2+\frac{\Ddot{a}}{a}\right],
\end{equation}
where
\begin{eqnarray}\label{eq11}
    \bar{M}(\psi)&=&\left(\frac{3F}{F'}\right)^2
    \frac{\partial^2\psi}{\partial r^2}+\frac{\partial^2\psi}{\partial \theta^2}+\frac{1}{\sin^2(\theta)}\frac{\partial^2\psi}{\partial \phi^2},\\
    \bar{N}(\psi)&=&\left(\frac{3F}{F'}\frac{\partial\psi}{\partial r}\right)^2+\left(\frac{\partial\psi}{\partial \theta}\right)^2+\frac{1}{\sin^2(\theta)}\left(\frac{\partial\psi}{\partial \phi}\right)^2,
\end{eqnarray}
and $H=\dot{a}/a$ is the local  Hubble parameter associated to the scale factor $a(t)$ of the collapsing metric. Hence, the Einstein equations result in the following system
\begin{equation}\label{eq13}
    \frac{2e^{2\psi}}{F^{2/3}a^2}\left[\bar{M}(\psi)-\bar{N}(\psi)+\left(\frac{12 F}{F'}-\frac{9 F''F^2}{F'^3}\right)\frac{\partial\psi}{\partial r}+\cot(\theta)\frac{\partial\psi}{\partial \theta}\right]+3e^{-2\psi}\left(H-\dot{\psi}\right)^2=4\pi G\left[\frac{e^{2\psi}\bar{N}(\varphi)}{a^2F^{2/3}}+\frac{\dot{\varphi}^2}{e^{2\psi}}+2V(\varphi)\right],
\end{equation}
\begin{equation}\label{eq14}
    e^{-2\psi}\left(5\dot{\psi}^2-8H\dot{\psi}-2\ddot{\psi}+H^2+2\frac{\ddot{a}}{a}\right)-\frac{e^{2\psi}}{F^{2/3}a^2}\bar{M}(\psi)=8\pi G\left\{\frac{e^{2\psi}}{6a^2F^{2/3}}\bar{N}(\varphi)-\frac{\dot{\varphi}^2}{2e^{2\psi}}+V(\varphi)\right\},
\end{equation}
\begin{equation}\label{eq15}
    \frac{\partial\psi}{\partial x^i}\frac{\partial\psi}{\partial x^j}=-4\pi G\frac{\partial\varphi}{\partial x^i}\frac{\partial\varphi}{\partial x^j},
\end{equation}
\begin{equation}\label{eq16}
    \frac{\partial}{\partial x^{i}}\left(\frac{\partial\psi}{\partial t}\right)+H\frac{\partial\psi}{\partial x^i}-\frac{\partial\psi}{\partial t}\frac{\partial\psi}{\partial x^i}=4\pi G \frac{\partial\varphi}{\partial t}\frac{\partial\varphi}{\partial x^i}.
\end{equation}
In addition, the dynamics of the inflaton field $\varphi$ is given by
\begin{equation}\label{eq17}
   \ddot{\varphi}+(3H-4\dot{\psi})\dot{\varphi}-\frac{e^{4\psi}}{a^2}\left[\frac{9}{F^{\prime}}\frac{\partial}{\partial r}\left(\frac{F^{4/3}}{F^{\prime}}\frac{\partial \varphi}{\partial r}\right)+\frac{1}{F^{2/3}sin\theta}\frac{\partial}{\partial\theta}\left(sin\theta\frac{\partial\varphi}{\partial\theta}\right)+\frac{1}{F^{2/3}sin^2\theta}\frac{\partial^2\varphi}{\partial\phi^2}\right]+e^{2\psi}V^{\prime}(\varphi)=0.
\end{equation}
Eliminating background terms and implementing an algebraic manipulation of equations (\ref{eq13}) and (\ref{eq14}) we obtain
\begin{eqnarray}\label{eq18}
&& e^{-2\psi}\left(\ddot{\psi}+7H\dot{\psi}-4\dot{\psi}^2\right)+\frac{e^{2\psi}}{a^2}\left[\frac{2}{3}\left(\frac{9F^{4/3}}{F^{\prime}\, ^{2}}\left(\frac{\partial\psi}{\partial r}\right)^2+\frac{1}{F^{2/3}}\left(\frac{\partial \psi}{\partial\theta}\right)^2+\frac{1}{F^{2/3}sin^2\theta}\left(\frac{\partial\psi}{\partial\phi}\right)^2\right)-\right.\nonumber\\
    && \left. \frac{9}{F^{\prime}}\frac{\partial}{\partial r}\left(\frac{F^{4/3}}{F^{\prime}}\frac{\partial\psi}{\partial r}\right)-\frac{1}{F^{2/3}sin\theta}\frac{\partial}{\partial\theta}\left(sin\theta\,\frac{\partial\psi}{\partial\theta}\right)-\frac{1}{F^{2/3}sin^2\theta}\frac{\partial^2\psi}{\partial\phi^2}\right]=-8\pi G\left[\frac{e^{2\psi}}{3a^2}\left(\frac{9F^{4/3}}{F^{\prime}\,^{2}}\left(\frac{\partial\varphi}{\partial r}\right)^2+\right.\right.\nonumber\\
    &&\left.\left. \frac{1}{F^{2/3}}\left(\frac{\partial\varphi}{\partial\theta}\right)^2+\frac{1}{F^{2/3}sin^2\theta}\left(\frac{\partial\varphi}{\partial\phi}\right)^2\right)+V(\varphi)\right].
\end{eqnarray}
Hence, the dynamics of $\psi$ is determined by equation \eqref{eq18}. In this manner a solution of \eqref{eq18} gives the form for gauge invariant scalar fluctuations of the metric characterized by an arbitrary amplitude. However, we will focus in this paper in the case of small fluctuations of quantum origin. 

\section{ The case of small amplitude fluctuations }

In general, determining solutions for the equation (\ref{eq18}) is not an easy task. However, for the particular case of small fluctuations viable solutions can be obtained. Small gauge invariant fluctuations of the metric can be study by means of  the weak field limit of the  equation \eqref{eq18}. Thus, at first-order, we can use the expansion formula
\begin{equation}\label{efor}
   e^{\pm n\psi}=1\pm n\psi. 
\end{equation}
 Let us also consider the semiclassical approximation for the inflaton field, in the form $\varphi(t,x^i)=\varphi_b(t)+\delta\varphi(t,x^i)$. Here, the background component of the field $\varphi$ is denoted as $\varphi_b=\langle E|\varphi|E\rangle$, which represents the expectation value of the field and $|E\rangle$ stands for a physical quantum state determined by the Bunch-Davies vacuum, and $\delta\varphi$ describes the quantum fluctuations of $\varphi$.\\

Thus, using the equations \eqref{eq16} and \eqref{efor} and considering the weak field limit in the equation (\ref{eq18})  we arrive to 
\begin{eqnarray}
    && \ddot{\psi}+\left(7H+\frac{2V'(\varphi_b)}{\dot{\varphi}_b}\right)\dot{\psi}-\frac{1}{a^2}\left[\frac{9}{F^{\prime}}\frac{\partial}{\partial r}\left(\frac{F^{4/3}}{F^{\prime}}\frac{\partial\psi}{\partial r}\right)+\frac{1}{F^{2/3}\,sin\theta}\frac{\partial}{\partial\theta}\left(sin\theta\,\frac{\partial\psi}{\partial\theta}\right)+\frac{1}{F^{2/3}sin^2\theta}\frac{\partial^2\psi}{\partial\phi^2}\right]+\nonumber\\
    &&
    2\left(3H^2+\dot{H}+\frac{V'(\varphi_b)}{\dot{\varphi}_b}H\right)\psi=0. \label{eq19}
\end{eqnarray}
On the other hand, the large scale background part coming from  the linearization of  the equation (\ref{eq17}) results to be
\begin{equation}\label{eq20}
    \ddot{\varphi}_b+3H_b\dot{\varphi}_b+V'(\varphi_b)=0,
\end{equation}
where $H_b=\left<H\right>$. At small quantum scales, the dynamics of $\delta\varphi$ is related to the dynamics of $\psi$ by means of the equation
\begin{eqnarray}
   && \delta\ddot{\varphi}+3H_b\delta\dot{\varphi}-4\dot{\psi}\delta\dot{\varphi}-\frac{1+4\psi}{a^2}
    \left[\frac{9}{F^{\prime}}\frac{\partial}{\partial r}\left(\frac{F^{4/3}}{F^{\prime}}\frac{\partial\delta\varphi}{\partial r}\right)+\frac{1}{F^{2/3}\,sin\theta}\frac{\partial}{\partial\theta}\left(sin\theta\,\frac{\partial\delta\varphi}{\partial\theta}\right)+\frac{1}{F^{2/3}sin^2\theta}\frac{\partial^2\delta\varphi}{\partial\phi^2}\right]+\nonumber\\
    &&(1+2\psi)V''\delta\varphi=4\dot{\psi}\dot{\varphi}_b-2\psi V'(\varphi_b).\label{eq21}
\end{eqnarray}
Now, assuming during inflation a slow-roll regime for the scalar field $\varphi$, the Friedmann equation for the background metric \eqref{eq6} reads
\begin{equation}\label{feq}
    H_b^{2}=\frac{8\pi G}{3}V(\varphi).
\end{equation}
Thus it follows from  (\ref{eq20}) that 
\begin{equation}\label{eq22}
    \dot{\varphi}_b^2=-\frac{\dot{H}_c}{4\pi G}.
\end{equation}
Hence, with the help of \eqref{feq} and  \eqref{eq22} the expression (\ref{eq19}) can be rewritten as
\begin{equation}
    \ddot{\psi}+H\dot{\psi}-\frac{1}{a^2}\left[\frac{9}{F^{\prime}}\frac{\partial}{\partial r}\left(\frac{F^{4/3}}{F^{\prime}}\frac{\partial\psi}{\partial r}\right)+\frac{1}{F^{2/3}\,sin\theta}\frac{\partial}{\partial\theta}\left(sin\theta\,\frac{\partial\psi}{\partial\theta}\right)+\frac{1}{F^{2/3}sin^2\theta}\frac{\partial^2\psi}{\partial\phi^2}\right] +2\dot{H}\psi=0.  \label{eq23}
\end{equation}
We are now in position to determine the power spectrum associated to the gauge invariant scalar fluctuations of the metric $\psi$. 

\section{The power spectrum}

In order to calculate the power spectrum for small scalar fluctuations of the metric we propose the separation $\psi(t,r,\theta,\phi)=\zeta(t,r)\Omega(\theta,\phi)$. Thus the equation \eqref{eq23} leaves to the system
\begin{eqnarray}\label{eq25}
    \frac{1}{\sin\theta}\frac{\partial}{\partial\theta}\left(\sin\theta\frac{\partial\Omega}{\partial\theta}\right)+\frac{1}{\sin^2\theta}\frac{\partial^2\Omega}{\partial\phi^2}+\ell(\ell+1)\Omega&=&0,\\
    \label{eq26}
    \ddot{\zeta}+H\dot{\zeta}-\frac{9}{a^2F'}\frac{\partial}{\partial r}\left(\frac{F^{4/3}}{F'}\frac{\partial\zeta}{\partial r}\right)+\left[2\dot{H}+\frac{\ell(\ell+1)}{a^2F^{2/3}
    }\right]\zeta&=&0,
\end{eqnarray}
where $\ell\in\mathbb{Z}^{+}\cup\lbrace 0\rbrace$ is a separation constant. The general solution of \eqref{eq25} is given in terms of the well known spherical harmonics $Y_{\ell,m}(\theta,\phi)$ with $m=-\ell...\ell$. Hence, the solution of \eqref{eq23} can be written in the form
\begin{equation}\label{solim}
\psi(t,\bar{r})=\psi(t,r,\theta,\phi)=\sum_{l=0}^{\infty}\sum_{m=-l}^{l}\zeta(t,r)Y_{l}^{m}(\theta,\phi) 
\end{equation}

Now, by means of the separation $\zeta(t, r) = T(t)P(r)$ the  equation (\ref{eq26}) gives
\begin{eqnarray}\label{eq27}
        \ddot{T}+H\dot{T}+2\dot{H}+\left(2\dot{H}+\frac{\alpha_t}{a^2}\right)T=0,\\
        \label{eq28}
        \frac{9}{F^{\prime}}\frac{d}{dr}\left(\frac{F^{4/3}}{F^{\prime}}\frac{dP}{dr}\right)+\left(\alpha_t-\frac{\ell(\ell+1)}{F^{2/3}}\right)P=0,
\end{eqnarray}
being $\alpha_t$ a separation constant. 
The general solution of (\ref{eq28}) reads
\begin{equation}\label{eq29}
        P(r)=\frac{C_1}{\sqrt{F^{1/3}}}J_{\ell+\frac{1}{2}}[\sqrt{\alpha_t}\,F^{1/3}]+\frac{C_2}{\sqrt{F^{1/3}}}Y_{\ell+\frac{1}{2}}[\sqrt{\alpha_t}\,F^{1/3}],
\end{equation}
where $C_1$ and $C_2$ are integration constants and $J_{\nu}$ and $Y_{\nu}$ are the first and second type Bessel functions. Now, by means of the transformation
\begin{equation}\label{eq30}
    \zeta(t,r)=e^{-\frac{1}{2}\int Hdt}\chi(t,r),
\end{equation}
the equation (\ref{eq26}) in terms of the auxiliary field $\chi$ becomes
\begin{equation}\label{eq31}
    \ddot{\chi}-\frac{9}{a^2F'}\frac{\partial}{\partial r}\left(\frac{F^{4/3}}{F'}\frac{\partial\chi}{\partial r}\right)+\left(\frac{\ell(\ell+1)}{a^2F^{2/3}}-\frac{H^2}{4}-\frac{3}{2}\dot{H}\right)\chi=0.
\end{equation}
Now, we consider the commutation relation
\begin{equation}\label{eq24}
    [\psi(\Bar{r}),\Pi^0_{(\psi)}(t,\Bar{r}')]=i r^2 sin\theta \, F^{\prime}\,e^{2\int H dt}\,\delta^{(3)}(\Bar{r}-\Bar{r}')
\end{equation}
where, the quantity $\Pi^0_{(\psi)}=\partial L/\partial\psi$ represents the canonical conjugate momentum of $\psi$ and $L$ is the Lagrangian, which is in this case is given by
\begin{equation}\label{lag}
    L=\sqrt{-g}\left[R+\frac{1}{2}g^{\alpha\beta}\varphi_{,\alpha}\varphi_{,\beta}-V(\varphi)\right].
\end{equation}
With the help of \eqref{eq10} and \eqref{efor} the relation \eqref{eq24} at first order in $\psi$ becomes
\begin{equation}\label{aeq1}
    \left[\psi(t,\bar{r}),\dot{\psi}(t,\bar{r})\right]=\frac{ir^2F^{\prime}\sin\theta \,e^{2\int H dt}}{36\sqrt{-g}}\,\delta^{(3)}\left(\bar{r}-\bar{r}^{\prime}\right).
\end{equation}
Thus, using the expression \eqref{solim} and the identities
\begin{eqnarray}\label{id1}
    && \delta^{(3)}\left(\bar{r}-\bar{r}^{\prime}\right) = \frac{1}{r^2 \sin\theta}\delta(r-r^{\prime})\delta(\theta-\theta^{\prime})\delta(\phi-\phi^{\prime}),\\
    && \sum_{l=0}^{\infty}\sum_{m=-l}^{l}Y_{l}^{m}(\theta,\phi)Y_{l}^{m}\,^{*}(\theta^{\prime},\phi^{\prime})=\frac{1}{\sin\theta}\delta(\theta-\theta^{\prime})\delta(\phi-\phi^{\prime}),
\end{eqnarray}
where the asterisck $*$ denotes complex conjugate, the relation \eqref{aeq1} yields
\begin{equation}\label{aeq2}
    \left[\zeta(t,r),\dot{\zeta}(t,r^{\prime})\right]=\frac{iF^{\prime}\sin\theta\,e^{2\int Hdt}}{36\sqrt{-g}} \delta(r-r^{\prime}).
\end{equation}
It follows from \eqref{eq6}  that $\sqrt{-g}=(1/3)a^3F^{\prime}\sin\theta$, therefore employing \eqref{eq8} the relation \eqref{aeq2} reduces to
\begin{equation}\label{aeq3}
    \left[\zeta(t,r),\dot{\zeta}(t,r^{\prime})\right]=\frac{i}{12 a_{\epsilon}^2 a}\,\delta(r-r^{\prime}),
\end{equation}
where $a_{\epsilon}=a(t_0+\epsilon t_0)$. Thus, with the help of \eqref{eq30} the expression \eqref{aeq3} reads
\begin{equation}\label{eq32}
    \left[\chi(t,\bar{r}),\dot{\chi}(t,\bar{r}')\right]=\frac{i}{12 a_\epsilon^3}\delta^{(3)}(\Bar{r}-\Bar{r}').
\end{equation}
On the other hand, we consider the Fourier expansion of $\chi$ in the form
\begin{equation}\label{eq33}
    \chi(\Bar{x},t)=\frac{1}{(2\pi)^{3/2}}\int d^3k_{r}\sum_{\ell=0}^{n-1}\left[\hat{a}_{k_{r}lm} P_{k_r}(r)\xi_{k_r}(t)+\hat{a}_{k_{r}lm}^\dagger P^*_{k_r}(r)\xi_{k_r}^{*}(t)\right],
\end{equation}
where for the dispersive case $\alpha_t=k_r^2$, we have
\begin{equation}\label{eq34}
    P_k(r)=\frac{C_1}{\sqrt{F^{1/3}}}J_{\ell+\frac{1}{2}}[k_rF^{1/3}]=C_{1}\sqrt{\frac{2 k_r}{\pi}}{\cal J}_{l}(k_r F^{1/3}),
\end{equation}
being ${\cal J}_l[z]$ the spherical Bessel function of the first type and where $\hat{a}_{k_{r}lm}^{\dagger}$ and $\hat{a}_{k_{r}lm}$ denote the creation and annihilation operators in spherical coordinates, respectively.  Thus we obtain for $\psi$ the expression
\begin{equation}\label{yqt1}
    \psi(t,r,\theta,\phi)=\frac{e^{-\frac{1}{2}\int H\,dt}}{(2\pi)^{3/2}}\int d^3k_r\sum_{l=0}^{n-1}\sum_{m=-l}^{l}\left[\hat{a}_{k_{r}lm}\,Y_{l}^{m}(\theta,\phi)P_{k_r}(r)\xi_{k_r}(t)+\hat{a}_{k_{r}lm}^{\dagger}\,Y_{l}^{m}\,^{*}(\theta,\phi)P_{k_r}^{*}(r)\xi_{k_r}^{*}(t)\right].
\end{equation}
Transforming \eqref{yqt1} to the cartesian coordinates $x=F^{1/3}sin\theta \,cos\theta$, $y=F^{1/3}sin\theta\, cos\theta$ and $z=F^{1/3} cos\phi$, we obtain 
\begin{equation}\label{yqt2}
    \psi(t,\bar{r})=\frac{e^{-\frac{1}{2}\int H\,dt}}{(2\pi)^{3/2}} \int d^3k \left[\hat{a}_{k}\,e^{i\bar{k}\cdot\bar{F}(\bar{r})^{1/3}}\xi_k(t)+\hat{a}_{k}^{\dagger}\,e^{i\bar{k}\cdot\bar{F}(\bar{r})^{1/3}}\xi_k^{*}(t)\right],
\end{equation}
where we have used the transformations for the creation and annihilation operators from spherical to cartesian representations 
\begin{eqnarray}\label{yqt3}
    \hat{a}_{k_{r}lm}^{\dagger}&=&(-i)^{l}k\int d\Omega(\theta,\phi)\,Y_{l}^{m}(\theta,\phi)\,\hat{a}_{k}^{\dagger},\\
    \label{yqt4}
    \hat{a}_{k_{r}lm}&=&i^{l}k\int d\Omega(\theta,\phi)\,Y_{l}^{m}\,^{*}(\theta,\phi)\, \hat{a}_k,
\end{eqnarray}
and the Rayleigh expansion 
\begin{equation} \label{yqt5}
    e^{i\bar{k}\cdot\bar{F}(\bar{r})^{1/3}}=4\pi\sum_{l=0}^{n-1}\sum_{m=-l}^{l}i^{l}{\cal J}_{l}(k_{r}F^{1/3})Y_{l}^{m}(\theta,\phi)Y_{l}^{m}\,^{*}(\theta,\phi).
\end{equation}
The creation-annihilation operators satisfy the algebra $\left[\hat{a}_k,\hat{a}_k^\dagger \right]=\delta^{(3)}(\bar{k}-\bar{k}')$, and $\left[\hat{a}_k,\hat{a}_{k'}\right]=\left[\hat{a}^\dagger_k,\hat{a}^\dagger_{k'}\right]=0$. Thus, the dynamics of the quantum modes $\xi_k$ for the scalar fluctuations of the metric is given by
\begin{equation}\label{eq35}
    \ddot{\xi}_k+\left[\frac{k^2}{a^2}-\left(\frac{H}{2}\right)^2+\frac{3}{2}\dot{H}\right]\xi_k=0.
\end{equation}
Now, in order to solve the equation \eqref{eq35}, we consider that the scale factor \eqref{eq7} can be expanded in the series
\begin{equation}\label{yq1}
    a(t)=\left(\frac{9}{4}\right)^{1/3}(t-t_0)^{2/3}+\frac{\Lambda}{3^{1/3}2^{8/3}}(t-t_0)^{8/3}+\cdots
\end{equation}
A representative term for each term in the series expansion \eqref{yq1} is given by
\begin{equation}
    \label{eq36}
    a(t)=\frac{3^{p-\frac{1}{3}}}{2^{2p-\frac{4}{3}}}\alpha_p\Lambda^{p-1}(t-t_0)^{2p-\frac{4}{3}},
\end{equation}
where $\lbrace\alpha_p\rbrace$ is a collection of constant parameters introduced to obtain the correct units and coefficients for each term in the expansion \eqref{yq1}. With the help of \eqref{eq36} the local Hubble parameter becomes 
\begin{equation}\label{emme}
H(t)=\frac{6p-4}{3(t-t_0)}.
\end{equation}
Using \eqref{eq36} and \eqref{emme} the equation \eqref{eq35} becomes
\begin{equation}\label{eq37}
    \ddot{\xi}_k+\left[\frac{2^{2(2p-\frac{4}{3})}k^2}{3^{2(p-\frac{1}{3})}\alpha_{p}^2\Lambda^{2(p-1)}(t-t_0)^{2(2p-\frac{4}{3})}}-\frac{p^2+\frac{5}{3}p-\frac{14}{9}}{(t-t_0)^2}\right]\xi_k=0.
\end{equation}
The general solution for \eqref{eq37} reads
\begin{equation}\label{eq38}
    \xi_{k}(t)=\delta_1\,\sqrt{t-t_0}\,\mathcal{H}^{(1)}_\nu[z(t)]+\delta_2\,\sqrt{t-t_0}\mathcal{H}^{(2)}_\nu[z(t)],
\end{equation}
where $\delta_1$, $\delta_2$ are integration constants and
\begin{eqnarray}\label{eq39}
    \nu&=&\frac{\sqrt{4p^2+\frac{20}{3}p+\frac{65}{9}}}{2\left(2p-\frac{7}{3}\right)},\\
    \label{eq40}
    z(t)&=&\frac{2^{2p-\frac{4}{3}}}{3^{p-\frac{1}{3}}\left(2p-\frac{1}{3}\right)\alpha_p\Lambda^{p-1}}\,k(t-t_0)^{\frac{7}{3}-2p}.
\end{eqnarray}
The functions $H_\nu^{(1)}$ and $H_\nu^{(2)}$ denote the first and second kind Hankel functions, respectively. With the help of \eqref{eq32} and \eqref{eq33} the normalization condition for the modes is given by
\begin{equation}\label{eq41}
    \xi_{k}(t)\dot{\xi}_{k}^*(t)-\xi_{k}^*(t)\dot{\xi}_{k}(t)=\frac{i}{12 a_\epsilon^3},
\end{equation}
where 
\begin{equation}\label{yab1}
    a_{\epsilon}=3^{p-\frac{1}{3}}\alpha_p \Lambda^{p-1}\left(\frac{\epsilon t_0}{2}\right)^{2p-\frac{4}{3}}.
\end{equation}
Using (\ref{eq41}) and choosing the Bunch-Davies vacuum the normalized solution for (\ref{eq37}) reads
\begin{equation}\label{eq 42}
    \xi_k(t)=\sqrt{\frac{\pi}{48a_{\epsilon}^3\left(\frac{7}{3}-2p\right)}}\,\sqrt{t-t_0}\,\mathcal{H}_\nu^{(1)}[z(t)].
\end{equation}
Now, the quadratic quantum fluctuations of $\psi$ in the infrared (IR) sector, at cosmological scales, are given by the expression
\begin{equation}\label{eq43}
    \langle\psi^2\rangle_{IR}=\frac{e^{-\int H dt}}{2\pi^2}\int^{\epsilon k_g}_0 \frac{dk}{k} k^3[\xi_k(t) \xi^*_k(t)]\Big|_{IR},
\end{equation}
where $k_g$ is the wave number related to the gravitational radius $r_g$. 
Therefore, the mean quadratic fluctuations for $\psi$ in the IR sector according to the equation (\ref{eq43}) are given by
\begin{equation}\label{eq44}
    \langle\psi^2\rangle=\frac{2^{6p-2\nu(2p-\frac{7}{3})-10}\Gamma^{2}(\nu)\left(2p-\frac{1}{3}\right)^{2\nu}\alpha_p^{-(3-2\nu)}(\epsilon t_0) ^{\frac{8}{3}-4p}}{(1-\nu)\pi^3 3^{\frac{5}{3}-2\nu (p-\frac{1}{3})}}\left(6p-4\right)^{(1-2\nu)\left(\frac{7}{3}-2p\right)}\Lambda^{(3-2\nu) (p-1)}H^{2p-\frac{4}{3}}\left(\epsilon k_g\right)^{2-2\nu}.
\end{equation}
The corresponding power spectrum for $\psi$ takes the form
\begin{equation}\label{eq45}
    \mathcal{P}_s(k)=\frac{2^{6p-2\nu(2p-\frac{7}{3})-9}\Gamma^{2}(\nu)\left(2p-\frac{1}{3}\right)^{2\nu}\alpha_p^{-(3-2\nu)}(\epsilon t_0) ^{\frac{8}{3}-4p}}{\pi^3 3^{\frac{5}{3}-2\nu (p-\frac{1}{3})}}\left(6p-4\right)^{(1-2\nu)\left(\frac{7}{3}-2p\right)}\Lambda^{(3-2\nu) (p-1)}H^{2p-\frac{4}{3}}\,k^{3-2\nu}.
\end{equation}
The spectral index is given by $n_s=4-2\nu$. Thus it is not difficult to verify that with the help of \eqref{eq39} we arrive to
\begin{equation}\label{eq46}
    n_s=4-\frac{\sqrt{4\left(p^2+\frac{5}{3}p-\frac{14}{9}\right)+1}}{\left(2p-\frac{4}{3}\right)-1}.
\end{equation}
According to the PLANCK observational data the spectral index ranges in the interval $n_s=0.965\pm 0.004$ \cite{PC}. Considering \eqref{eq46}, these values for $n_s$ can be obtained when $p\in [1.958904, 1.962291]$. Therefore, we can say that a nearly scale invariant spectrum can be obtained when $p\simeq 2$. For the observational values of $n_s$ it corresponds $\nu=1.525\mp 0.002$. Thus we can express $\nu$ as: $\nu=\frac{3}{2}+\delta\nu$ with $\delta\nu=0.025\mp 0.002$. In view of \eqref{eq45} the nearly scale invariant spectrum has approximately the form
\begin{equation}\label{qnal1}
{\cal P}_{s}\sim \alpha_p^{2\delta\nu}\Lambda^{2\delta\nu}H^{\frac{8}{3}}
\end{equation}

On the other hand, three independent missions have now confirmed the lack of angular correlation in the CMB temperature profile at angles $\gtrsim 60^{\circ}$ \cite{Chal}. One likely explanation for this anomaly is the existence of a cutoff in the primordial power spectrum given by  $k_{min}=(3.14\pm 0.36)\times 10^{-4}\,Mpc^{-1}$. One possible implication of this cutoff is that there exists a specific time, beyond the Planck time, at which the inflationary modes started exiting the horizon \cite{Chal}.  Thus, with the effect of producing additional degrees of freedom an option is to assume a pre-inflationary era that begins at the Planck's time. However, it is important to remember that one of the requirements for an inflationary model to be viable is that it must produce a sufficient number of e-foldings to solve the horizon problem. However, if the inflationary suffers a time reduction because of the cutoff, so it does the number of e-foldings. Thus, the idea is that the pre-inflationary period contributes to the missing e-foldings. If it were the case where the pre-inflationary model did not provide enough e-foldings  to solve the horizon problem,  the inflationary period would have to end later that expected and this would imply that the amplitude of the power spectrum would decrease, leaving to another non-correspondence with observational data. However, one possibility to recover the amplitude of the power spectrum in this hypothetical case is to consider the spectrum generated by the formation of scalar black holes at the end of the inflationary period as a correction to usual power spectrum generated by the inflaton fluctuations at the end of inflation.  In this manner, with the help of \eqref{qnal1}, the total amplitude can be approximated by
\begin{equation}\label{qnal2}
    A_{T}\simeq \alpha_p^{2\delta\nu}\Lambda^{2\delta\nu}H^{\frac{8}{3}}+H_{end}^2=H_{63}^2,
\end{equation}
where $H$ is the Hubble parameter associated to the collapsed metric \eqref{eq6}, $H_{end}$ is the Hubble parameter given by the inflationary model considered after the pre-inflationary period, and $H_{63}$ is the Hubble parameter given after the $63$ e-foldings enough to solve the horizon problem. Notice that the first term is the correction coming from the generation of scalar black holes and is depending of the local vacuum at the moment of the formation of a scalar black hole given by $\Lambda$. Hence, it follows from \eqref{qnal2} that the value of $\Lambda$ required to the correction to be enough to solve the horizon problem is given by
\begin{equation}\label{qnal3}
    \Lambda=\frac{(H_{63}^2-H_{end}^2)^{1/(2\delta\nu)}}{\alpha_p H^{4/(3\delta\nu)}}.
\end{equation}
On the other hand, according to \cite{Carneiro}, the mass of the scalar black holes generated is given by the formula
\begin{equation}\label{qnal4}
    M=\frac{4\pi}{\sqrt{\Lambda}}.
\end{equation}
Thus, the mass of the scalar black holes generated to have the local vacuum value \eqref{qnal4} results to be 
\begin{equation}\label{qnal5}
    M=\frac{4\pi \,\alpha_p^{1/2}H^{2/(3\delta\nu)}}{(H_{63}^2-H_{end}^2)^{1/(4\delta\nu)}}.
\end{equation}
Therefore, we can interpret the fact that the power spectrum generated during the formation of a scalar black hole at the end of inflation could be nearly scale invariant, opens the possibility to add a correction to the power spectrum generated by the inflaton quantum fluctuations  during inflation.

\section{Final comments}

In this paper we have calculated the power spectrum of gauge invariant fluctuations of the metric generated during the local collapse of vacuum determined by the local inflaton scalar field,  near the end of inflation. We have employed the Carneiro and Fabris collapse mechanism based in the separation of the local  energy-momentum tensor into a pressureless and a vacuum parts. Considering that only the vacuum component contributes, we have calculated the power spectrum associated to the collapse metric  fluctuations in the weak field limit. By considering a series expansion of the scale factor and considering a representative term, we obtain the nearly scale invariance of the spectrum approximately for $p\simeq 2$.  The fact of recovering a nearly scale invariance for a second orden approximation in the scale factor, opens the possibility to consider the amplitude of the spectrum generated by the foprmation of scalar black holes as a correction to the total amplitude of the power spectrum generated by quantum inflaton fluctuations at the end of inflation. Finally, in view of the lack of angular correlation in the CMB  temperature profile at angles $\gtrsim 60^{\circ}$ \cite{Chal}, we have considered the situation in which a inflationary period after a pre-inflationary period do not generate the sufficient e-foldings to solve the horizon problem, and we have shown that in roder to the correction coming from the formation of scalar black holes would be enough to solve the horizon problem, the mass of the scalar black holes formed must be given by \eqref{qnal5}.

\section*{Acknowledgements}

\noindent J. E. Madriz-Aguilar, J. O. Valle and M. Montes acknowledge  CONACYT
(M\'exico) and Centro Universitario de Ciencias Exactas e Ingenierias of Guadalajara University for financial support. C. Romero thanks CNPq (Brazil) for financial support.





\bigskip

\begin{thebibliography}{99}

\bibitem{RI1} S. Hawking, Mon. Not. Roy. Astron. Soc. {\bf 152} (1971) 75.
\bibitem{RI2} B. J. Carr, S. W. Hawking, Mon. Not. Roy. Astron. Soc. {\bf 168} (1974) 399.
\bibitem{RI3} I. Stamov, S. Clesse, ``Can primordial black holes form in the standard model?", arXiv: 2312.06873v1 [astro-ph.CO] (2023).
\bibitem{RI4} S. G. Rubin, M. Y. Khlopov and A. S. Sakharov, Grav. Cosmol. {\bf 6} (2000) 51. ArXiv: hep-ph/0005271.
\bibitem{RI5} S. G. Rubin, A. S. Sakharov and M. Y. Khopov. J. Exp. Theor. Phys. {\bf 91} (2001) 921. ArXiv:hep-ph/0106187.
\bibitem{RI6} S. Pi Y., L. Zhang, Q. G. Huang, M. Sasaki, JCAP {\bf 05} (2018) 042. ArXiv: 1712.09896v2/astro-ph.CO.
\bibitem{RI7} J. Yokoyama, Astron. Astrophys. {\bf 318} (1997) 673. ArXiv: astro-ph.CO/9509027
\bibitem{RI8} E. Cotner, A. Kusenko, M. Sasaki, V. Takhistov. JCAP 10 (2019) 077.
\bibitem{RI9} S. G. Rubin, M. Y. Khlopov and A. S. Sakharov, Grav. Cosmol. {\bf 6} (2000) 51. ArXiv: hep-ph/0005271.

\bibitem{Carneiro} S. Carneiro, J. C. Fabris, Eur. Phys. J. C (2018) 78, 676.
\bibitem{Anabitarte} M. Anabitarte, M. Bellini, J. E. Madriz-Aguilar, Eur. Phys. J. C {\bf 65} (2010) 295-301.
\bibitem{Exp1} G. Hinshaw, A. J. Branday, C. L. Bennet, et al., The Astrophy. J. 464, (1996) L25-L28. 
\bibitem{Exp2} C. L. Bennet, R. S. Hill, G. Hinshaw, et al., The Astrophy. J. Supp. Series 148 (2003) 97-117.
\bibitem{Exp3} N.  Aghanim, Y. Akrami, et al. A\&A, 641 (2020) A6.
\bibitem{PC} N. Aghanim et al., (Planck Collaboration), Planck 2018 results. ArXiv:1807.06209v4[astro-ph.CO].
\bibitem{Chal} J. Liu and F. Melia, Astrophys. J. {\bf 967} (2024) 109.



\end{thebibliography}
\end{document}